\documentstyle[epsfig]{aipproc}

\newcommand{\al}{\mbox{$\alpha$}}
 
 \newcommand{\th}{\mbox{$\theta$}}
\newcommand {\mod}[1] {\mbox{$ \vert #1 \vert $}}
 \newcommand{\ga}{\mbox{$\gamma$}}
 
\newcommand{\Ga}{\mbox{$\Gamma$}}

\newcommand{\bec}[1] {\begin{equation}\label{#1} }
\newcommand{\eec} {\end{equation} }
 
\newcommand{\ti}[1]{\mbox{$ \tilde{#1} $} }

\begin{document}

\begin{flushright}
UR-1542\\
ER/40685/921\\
hep-ph/9808403\\
August 1998 \\
\end{flushright}

\title{Gluon Radiation \\ in Top Quark Production and Decay \\
 at an $e^+ e^-$ Collider\thanks{Presented by C.~Macesanu
  at 20th Annual MRST (Montreal-Rochester-Syracuse-Toronto) Meeting 
  on High-Energy Physics, Montreal, 13-15 May 1998. } }

\author{Cosmin Macesanu and Lynne H. Orr}
\address{Department of Physics and Astronomy,
University of Rochester\\
Rochester, New York 14627-0171}

\maketitle

\begin{abstract}
We study the effects of gluon radiation on top production and decay 
processes at an $e^+e^-$ collider. The matrix elements are computed without 
any approximations, using spinor techniques. We use a Monte 
Carlo event generator which takes into account the infrared singularity
due to soft gluons  and differences in kinematics
associated with radiation in the production versus decay process. 
The calculation is illustrated for several strategies of top mass
reconstruction.
\end{abstract}

\section*{Introduction}

 The study of the top quark is one of the most important goals for 
future high energy experiments. Top 
has several characteristics which set it apart from the other quarks,
the most obvious being 
its large mass:  at 175 GeV  top  is 
35 times heavier than its partner the $b$.
The large mass of the top quark has a number of interesting consequences.  It 
makes $t\bar{t}$ loop corrections to electroweak processes important,
so that  a good measurement
of $m_t$ coupled with  extremely precise values of the $W$ mass
will give limits on the  Higgs mass.
 A heavy top may
decay into  supersymmetric (SUSY) particles, providing a way of testing 
predictions of SUSY models.
Perhaps most important,   
large mass  means that  
the top Yukawa coupling to the Higgs is large; actually it is 
close to unity. This means that top
studies  can offer insight into  electroweak symmetry breaking, 
as well as into the fermion mass generation process.

 Another consequence of the top quark's large mass is that 
 its  decay width is about 1.5 GeV, which means  its lifetime  
is of order $10^{-24} $ seconds. In this short a time, the top quark
doesn't have time to hadronize. As a consequence, 
  the top can be completely described by perturbative QCD. Also, 
 the  spin information  is transmitted to the decay products 
($b$ and a $W^+$). 
This allows us  to get information about the top couplings from angular 
distributions of the resulting particles; 
differences between measured  and  SM values 
would  indicate new physics. 
 
The top was discovered in 1995 at the Fermilab Tevatron proton-antiproton
collider.   This machine is now being upgraded, and in 2000 the study of the
top here will start again. Also, at the Large Hadron Collider precise 
measurements  of the top parameters (mass, couplings, production
cross-section) will be one of the priorities. 

Besides studies at these machines, it would be interesting to 
study top at a  high-energy electron-positron collider.  
 The principal advantage of a 
 lepton collider versus a hadronic collider is a much cleaner environment;
 the task of performing certain precision measurements will be much easier in
 the absence of large QCD backgrounds.   In particular, studies
 at the $t \bar t$ threshold can be done at a lepton collider, but such
 studies are impossible at a hadron collider.
 Measurements of top quark couplings are especially difficult at a hadron 
collider. At an $e^+e^-$ collider, however, besides smaller 
background, the possibility of using
 polarized electrons in the 
initial state will be an advantage; also, the coupling of the 
top to neutral currents ($\ga$ and $Z$) will be accessible.  
In general, the capabilities of a  high energy lepton collider 
are complementary to those of a hadron collider.
   
To be able to perform these studies, though, a good theoretical 
understanding of top quark's production and decay processes is
essential.  
An important issue is QCD corrections due to real or virtual
gluon emission because of the large value
 of the strong coupling constant ($\approx 0.1$). 
In 
 the following, we will be concerned with 
 corrections due to the radiation of a real gluon.

\section*{Calculational Procedure} 
   
We are interested in  top production and decay with emission of a gluon:
$$ e^+e^- \rightarrow \ga^*, Z^* \rightarrow t\bar{t} (g)
\rightarrow bW^+ \bar{b}W^-(g)$$ 
(the top decays into a $b- W$ pair with a branching ratio close to 
unity).
The  gluon radiation  can take place  during the 
$t\bar{t}$ production process or during the decay of either the $t$ or 
$\bar{t}$.
Studies of this subject have focused on
a single portion of the entire process (such as corrections to production
\cite{mcos:jersak,mcos:doc} or decay \cite{mcos:jeza} only), or 
imposed approximations
such as soft gluons  \cite{mcos:kos} or intermediate on-shell top quarks and
massless $b$ quarks \cite{mcos:schmidt}.
The results presented here are obtained
using an exact computation of the entire matrix element for real
gluon emission, including 
all spin correlations, top width, and $b$ mass effects\cite{mcos:macorr}.

   \begin{figure}[t!] 
\centerline{\epsfig{file=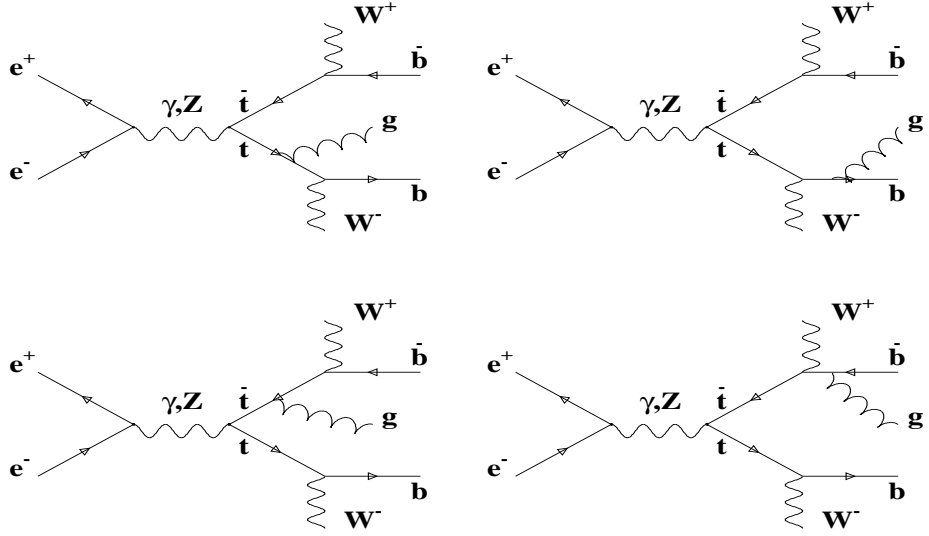,height=3.5in,width=4.5in}}
\vspace{10pt}
\caption{  Feynman diagrams for top quark production with gluon radiation}
\label{mcos:fig1}
\end{figure}

  Let  us briefly describe the procedure for obtaining these results.
We start by evaluating the amplitude for the process; there are four 
diagrams which contribute, (see Fig. \ref{mcos:fig1}), so we will have:
\bec{mcos:a11}
  M= \frac{B}{P_{tg}*P_{\bar{t}}} + \frac{BB}{P_{t}*P_{\bar{t}g}} + 
\frac{T}{P_{t}*P_{tg}*P_{\bar{t}}} + \frac{TB}{P_{t}*P_{\bar{t}}*P_{\bar{t}g}}
\eec
with
\bec{mcos:a12}
  P_{t}=p_t ^2 -m_t ^2 + i m_t \Ga _t \ ; \ p_t=p_{W^+}+p_b
\eec
$$ P_{tg}=p_{tg} ^2 -m_t ^2 + i m_t \Ga _t \ ; \ 
p_{tg}=p_{W^+}+p_b+p_g   $$
and similar definitions for the denominators of $\bar{t}$ propagators.

  The $B, \ BB$ terms correspond to the gluon being radiated by
the $b, \ \bar{b}$ quarks.
The last two terms correspond to a gluon radiated by the $t$ or $\bar{t}$,
 either in the production or decay stage. To separate these cases, we 
rewrite the propagator products as follows:

$$ \frac{1}{P_{t}*P_{tg}}=\frac{1}{2p_g p_t} [ \frac{1}{P_{t}}-\frac{1}{P_{tg}} ] \ , \
 \frac{1}{P_{\bar{t}}*P_{\bar{t}g}}=\frac{1}{2p_g p_{\bar{t}} } [ \frac{1}{P_{\bar{t}}}-
\frac{1}{P_{\bar{t}g}} ]$$   
and the amplitude can be writen 
\bec{mcos:a13}
 M=A1+A2+A3
\eec
with
$$ A1= \frac{1}{P_{tg}*P_{\bar{t}}} [B- \frac{T}{2p_g p_t}]
\ , \
 A2= \frac{1}{P_{t}*P_{\bar{t}g}} [BB- \frac{TB}{2p_g p_{\bar{t}}}] 
 $$ 
$$ A3= \frac{1}{P_{t}*P_{\bar{t}}} [ \frac{T}{2p_g p_t} + 
 \frac{TB}{2p_g p_{\bar{t}}}]   $$
  We identify these three terms as corresponding to gluon 
radiation in the $t$ decay ($A1$), $\bar{t}$ decay ($A2$) or production stage
($A3$). 

The three separate parts of the amplitude are calculated using helicity 
amplitudes and can be evaluated
 numerically.
We can then compute each of the six resulting terms
  separately:
   $$ \sum_{helicities} \mod{M}^2 = 
  \sum_{helicities} \left\{ \mod{A1}^2+\mod{A2}^2+\mod{A3}^2+
  2Re[ A1 A2^* + A1 A3^* + A2 A3^*] \right\}$$
 
To integrate this formula over the phase space
we use a three-channel Monte Carlo with one channel for each 
of the diagonal terms $\mod{A1}^2,\mod{A2}^2,\mod{A3}^2$   and a combination 
 of channels for the interference terms $ 2Re[ A1 A2^* + A1 A3^* + A2 A3^*]$.
The phase space region where the gluon energy $E_g$ goes to $0$ presents some
problems, though, because the amplitude has a singularity there. Even with 
cuts on $E_g$, the rapid variation of the integrand can spoil the integration
procedure. To eliminate this problem, we tailor the momentum generator
to the production of a gluon in association with two massive particles
($\ga^*, Z^* \rightarrow t \bar{t} g$ or $t  \rightarrow b W g $).
    
 \section*{Numerical Results}

 In this section, we  present some preliminary results obtained using 
the procedure described above. As input, we use the following values:
$m_t=175$ GeV, $\Ga _t=1.5$ GeV,
 $m_b=5$ GeV, 
$\al _s=0.1$, and a center of mass energy $W=600$ GeV. Also, unless
otherwise specified, we impose a cut on gluon energy $E_g>10$ GeV. 

 First, we note  that the radiation of a gluon
plays an important role in $t \bar{t}$ production. The lowest-order, tree-level
cross section  
is $\sigma_0 =0.43\ {\rm pb}$, 
while the cross section for the process with emission of a gluon with energy 
greater than 10 GeV 
is $\sigma_0 =0.39\ {\rm pb}$.\footnote{Note that this number includes 
contributions
from decay-stage radiation which are not corrections to the total
top production cross section.} 
We see that the two quantities have the same
order of magnitude.  

 \begin{figure}[t!] 
\centerline{\epsfig{file=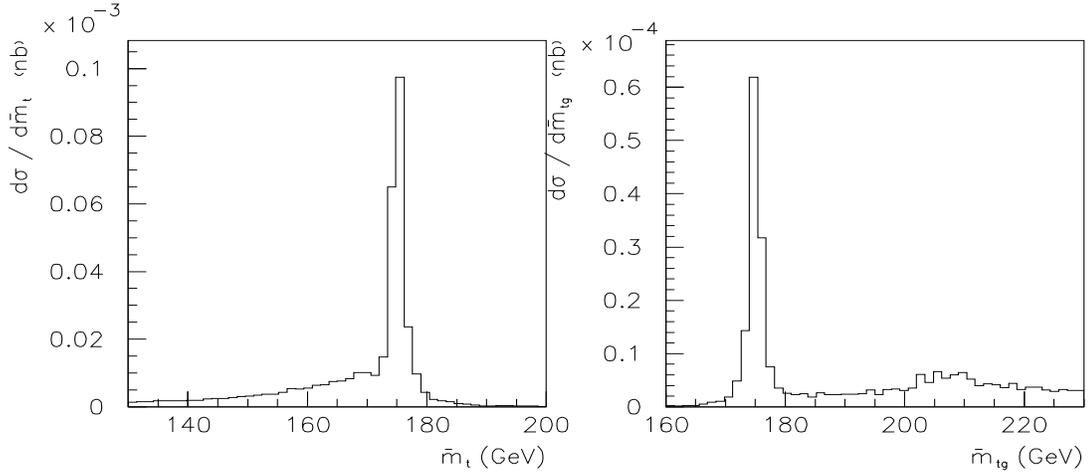,height=3.0in,width=4.5in}}
\vspace{10pt}
\caption{Distributions in  
  $\tilde{m}_t$  and $\tilde{m}_{tg}$ as defined in the text.}
\label{mcos:fig2}
\end{figure}
\begin{figure}[b!] 
\centerline{\epsfig{file=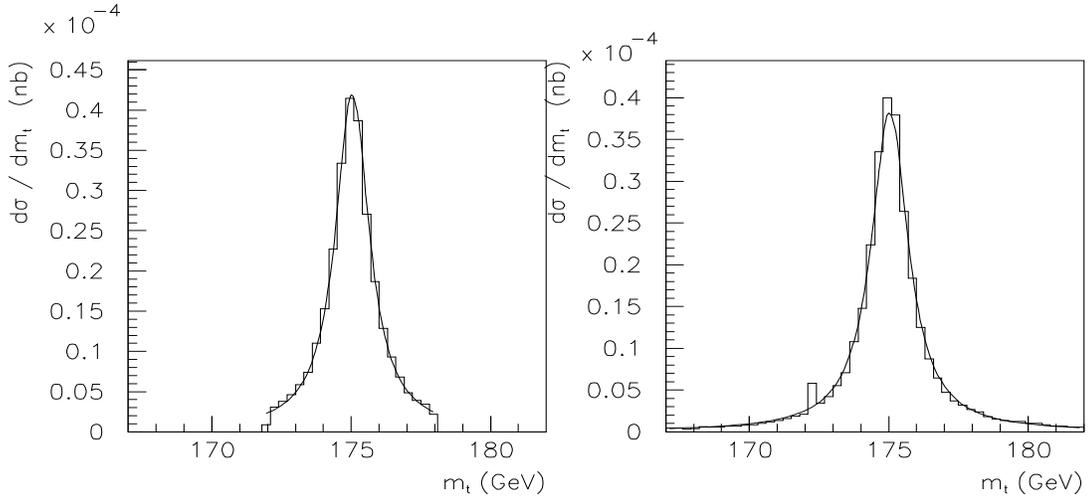,height=3.0in,width=4.5in}}
\vspace{10pt}
\caption{Mass distributions using mass cuts (left) and gluon angle cuts
(right).
 }
\label{mcos:fig3}
\end{figure}      
   
  An interesting (and important) issue is top mass reconstruction.
   This is closely 
related to the problem of assigning the gluon to the correct process in which
it was radiated. For example, if the gluon was radiated in the 
production stage,
the mass of the top is given by  
\bec{mcos:c1}
m_t^2=p_t^2=(p_b+p_{W^+})^2 \eec
 On the
other hand, if the gluon was radiated in $t$ decay stage, the above formula 
will give us a low estimate of $m_t$; the correct formula will be 
\bec{mcos:c2}
m_t^2=p_{tg}^2=(p_b+p_{W^+} +p_g)^2 \eec
 In Fig. \ref{mcos:fig2} we present 
the mass distributions obtained using (\ref{mcos:c1}) and (\ref{mcos:c2})
 respectively.
While there are clear peaks at the input values of the top mass, we see 
significant tails due to wrong gluon assignments.  The existence of 
such tails can increase measurement uncertainties on the top mass
or confound attempts to identify top events by mass reconstruction.
It is apparent that, for good mass reconstruction,
 we need to find a method to assign the gluon to the
correct process in which it was radiated.

 Using the variables $\ti{m}_t=\sqrt{p_t^2}$, 
 $\ti{m}_{tg}=\sqrt{p_{tg}^2}$, $\ti{m}_{\bar t}=\sqrt{p_{\bar t}^2}$ and
 $\ti{m}_{{\bar t}g}=\sqrt{p_{{\bar t}g}^2}$  
  we define four types of events: 

\hspace{0.5in} -type 1 : $172\ GeV < \ti{m}_{tg} , \ti{m}_{\bar{t}} <178\ GeV$ 

\hspace{0.5in} -type 2 : $172\ GeV < \ti{m}_{t} , \ti{m}_{\bar{t}g} <178\ GeV$ 

\hspace{0.5in} -type 3 : $172\ GeV < \ti{m}_{t} , \ti{m}_{\bar{t}} <178\ GeV$ 

    \hspace{0.5in} -type 4 : any other event  

  \begin{figure}[b!] 
\centerline{\epsfig{file=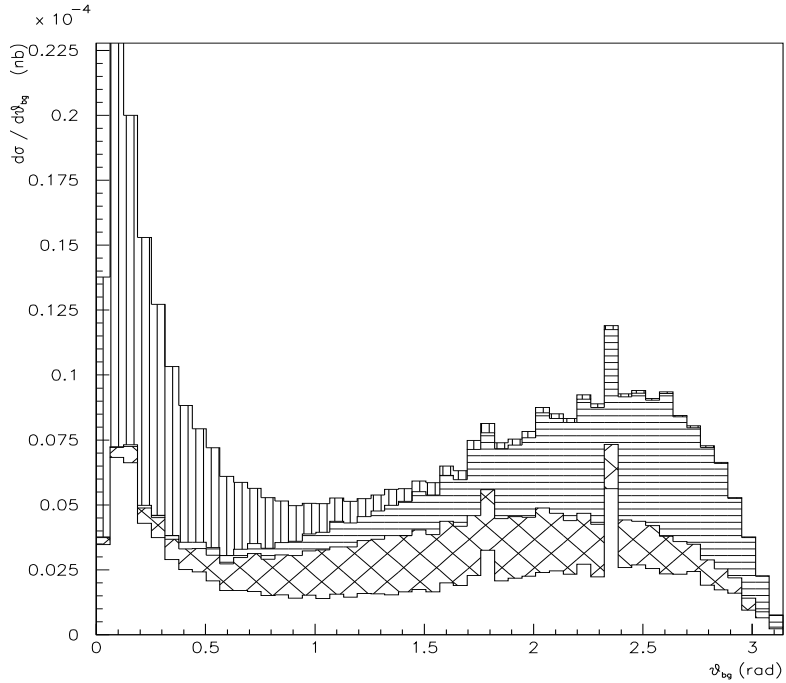,height=3.5in,width=3.5in}}
\vspace{10pt}
\caption{ gluon-b angle distribution}
\label{mcos:fig4}
\end{figure}         
    
We identify type 1 events as corresponding to gluons radiated in $t$ decay
stage. In this case we expect $p_{\bar t}$, $p_{tg}$ to be on shell; 
however, as we use a Breit-Wigner distribution, we  accept deviations
from the exact value $m_t=175$ GeV by about $2\Ga _t =3$ GeV. Furthermore,
we identify    
 type 2 events as corresponding to gluons radiated in $\bar{t}$ decay
stage and type 3  as corresponding to gluons radiated in $t \bar{t}$ production
stage. As the gluon energy is quite big ($>10$ GeV) compared with
$\Ga _t$, we expect this interpretation to work; meaning that there will be 
very few events which satisfy one of the conditions 1, 2, 3 but which
are open to more than one interpretation. Finally, 
type 4 would correspond  to events for which there is no compelling 
evidence for either the production- or decay-stage radiation case; 
it will actually be a mixing of both. 

In the first plot Fig. \ref{mcos:fig3} 
we present the top mass distribution using events of types 1, 2,
and 3 defined above.
We have  generated 250,000 events for each case. The smooth line 
is a Breit-Wigner function fitted over the mass distribution. The fit 
reproduces the input parameters with remarkable accuracy; it actually 
gives us 175.04 GeV for $m_t$ and 
1.46 GeV for $\Ga _t$.

  Another method for reconstructing the process makes use of the gluon-$b$
and gluon-$\bar{b}$ angle distributions. It is known that a gluon radiated
by a quark tends to go in the same direction as that quark.\footnote{Since
we keep the $b$ mass, there is strictly speaking no collinear singularity;
however gluons tend to be radiated from the $b$ at small angles 
nonetheless.}
Then, the gluons
close to the $b$ quark probably came from the top decay process; similarly,
the gluons close to the $\bar{b}$ are likely to have come from $\bar{t}$
decay. We might guess  that the rest of the gluons may be assigned to the 
production process.

In Fig. \ref{mcos:fig4} we present the 
distribution of the angle $\th _{bg}$ between
the gluon and the $b$ quark. The vertical-hatched part corresponds 
to type 1 events; and they are close to the $b$ direction. The 
horizontal-hatched part 
corresponds to type 2 events; as they will cluster near the
$\bar{b}$ quark, and $b$ and $\bar{b}$ are mostly in opposite hemispheres, 
these 
events tend to gather at large angles. The cross-hatched part corresponds
to type 3 events, and they are distributed uniformly. Finally, the
non-hatched part corresponds to type 4 events, and, as they are an admixture
of the first three, it looks like the whole distribution on a smaller scale.

 Using this figure we can make the following conventions:
 
 \hspace{0.5in} -if $\th _{bg} < 0.7 \ rad $ assign gluon to $t$ decay 
 
 \hspace{0.5in} -if $\th _{\bar{b}g} < 0.7 \ rad$ assign gluon to $\bar{t}$
  decay 
  
 \hspace{0.5in} -if $\th _{bg}, \th _{\bar{b}g} >1\ rad $ 
 assign gluon to $t\bar{t}$ production.
 
With these definitions, we construct  the top mass distribution  presented
in the second plot in Fig. \ref{mcos:fig3}.
Again, a fit with a Breit-Wigner function gives us 
values very close to our input parameters:  
$m_t=175.05 $ GeV  , $ \Ga_t =1.6$ GeV.

\begin{figure}[t!] 
\centerline{\epsfig{file=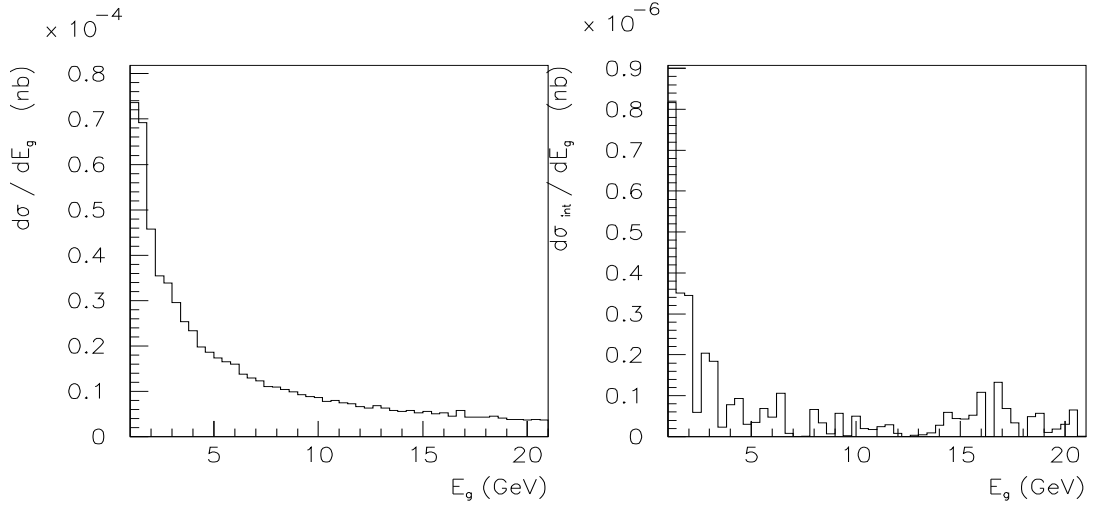,height=3.0in,width=4.5in}}
\vspace{10pt}
\caption{Gluon energy distribution; total cross-section and 
interference contribution, respectively.}
\label{mcos:fig5}
\end{figure} 
        
  \begin{figure}[t!] 
\centerline{\epsfig{file=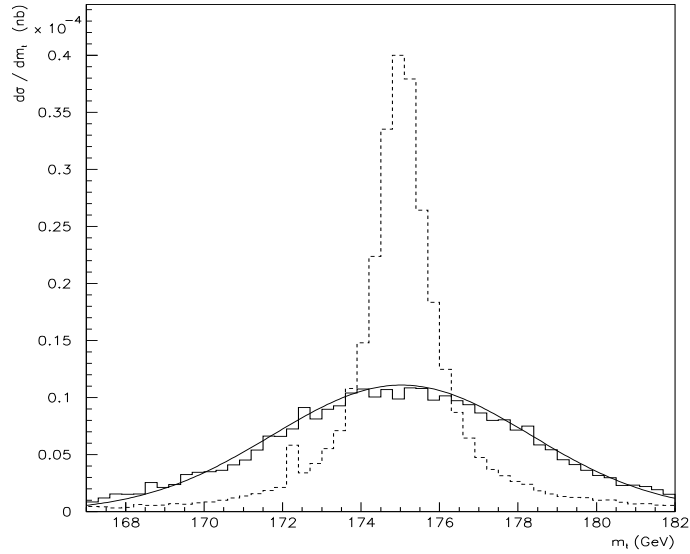,height=3.0in,width=3.0in}}
\vspace{10pt}
\caption{Mass reconstruction distribution, taking into account detector 
resolution (the dashed line corresponds to exact energy values).}
\label{mcos:fig6}
\end{figure}

Fig. \ref{mcos:fig5} presents the gluon energy distribution. The soft 
gluon singularity is visible (we have used a $E_g>1$ GeV  cut).
 This raises an interesting problem: soft gluons are not visible 
 in the detector; therefore, the process in which  a soft gluon is
 radiated will give the same signal as the process in which there is 
 no gluon. The study of the  experimental cross section for this last type of 
 process will have to take into account radiation of gluons with energies 
 up to $5$ GeV, maybe.  This also requires including virtual corrections, 
as in \cite{mcos:schmidt}.
 
  At this point, it is 
interesting to see if the interference terms contribute.  We can 
integrate separately the contribution of diagonal terms and the contribution
of interference terms; it turns out that this last contribution amounts 
only to around 1$\%$ of the total distribution, as shown in Fig. \ref{mcos:fig5}.

 Up to here, we have worked at the parton level and 
 presumed exact information about the momenta of particles 
in the final state. In a real experiment, though, there are a lot of 
complications: 
the  electrons in the initial state can lose energy through radiation
before interacting, thus having a lower CM energy; we don't see neutrinos,
don't know which jet corresponds to which quark (or gluon), and the 
energies measured
are not exact.  One can expect that the detector resolution 
will have one of the biggest effect on the top mass (and width) reconstruction.

 In Fig. \ref{mcos:fig6} we present the distribution for the top mass
(using gluon angle cuts) obtained after taking into account this effect.
The spread in the measured energies is parametrized by gaussians with
widths $\sigma=0.4 \sqrt{E}$ for quarks and gluon, and $ 
\sigma=0.15 \sqrt{E}$ for the $W$'s. The $m_t$ value obtained in 
this case is still very good ($175.03$ GeV) but the width of the 
distribution in this case reflects rather the detector resolution
effects than the top width.
 
\section*{Conclusions}
We have presented preliminary results of a calculation of real gluon
radiation in top production and decay.\cite{mcos:macorr} 
 Mass reconstruction using various mass cuts and gluon-bottom quark angle cuts
was performed. In an ideal situation, (no initial state radiation, 
perfect particle identification, exact energy and angle measurements)
the results were found to be  very good, {\it i.e.}, the distributions
faithfully reproduced the input mass and width.    
In a further study, we will have to take into account the effects of 
energy smearing due to detector resolution, initial state radiation, 
undetected neutrinos and jet identification problems.
Effects due to interference between gluon radiation in the production
and decay stages do not appear to be experimentally visible, at least for
hard gluons. Soft gluon radiation (the experimental
signal in this case will mimic that of a process without any gluon) and 
virtual gluon corrections have to be studied  in more detail.

We thank C.R.~Schmidt and W.J.~Stirling for helpful correspondence and 
discussions.  This work was supported in part by the U.S.~Department of Energy
and the National Science Foundation.

\end{document}